\documentclass[twocolumn,aps,showpacs,superscriptaddress]{revtex4-1}
\usepackage{amsmath}
\usepackage{graphicx}
\usepackage{slashed}
\RequirePackage[pagewise,mathlines]{lineno}

\begin{document}
%\linenumbers
\title{Exploring the double-slit interference with linearly polarized photons}%
\author{Wangmei Zha}\affiliation{University of Science and Technology of China, Hefei, China}
\author{James Daniel Brandenburg}\email{dbrandenburg.ufl@gmail.com}\affiliation{Shandong University, Jinan, China}\affiliation{Brookhaven National Laboratory, New York, USA}

\author{Lijuan Ruan}\affiliation{Brookhaven National Laboratory, New York, USA}
\author{Zebo Tang}\email{zbtang@ustc.edu.cn}\affiliation{University of Science and Technology of China, Hefei, China}
\author{Zhangbu Xu}\affiliation{Brookhaven National Laboratory, New York, USA}\affiliation{Shandong University, Jinan, China}

%\author{The American Physical Society}%
%\email[REVTeX Support: ]{revtex@aps.org}
%\affiliation{1 Research Road, Ridge, NY 11961}
\date{\today}%
\begin{abstract}
The linearly polarized quasi-real photons from the highly Lorentz-contracted Coulomb fields of relativistic heavy ions can fluctuate to quark-antiquark pairs, scatter off a target nucleus and emerge as vector mesons. In the process, the two colliding nuclei can switch roles to act as photon emitter or target, forming a double-slit interference pattern. The product from photoproduction inherits the photon polarization states, leading to the asymmetries of the decay angular distributions. In this letter, we study the interference effect in polarization dimension from the asymmetries of the decay angular distributions for photoprodution in heavy-ion collisions and find a periodic oscillation with the transverse momentum of vector meson, which could reasonably explain the transverse momentum dependence of the 2nd-order modulation in azimuth for the $\rho^{0}$ decay observed by the STAR collaboration.  
\end{abstract}
\maketitle
In relativistic heavy-ion collisions, the giant electromagnetic field accompanied by the heavy nuclei can be viewed as a spectrum of quasi-real photons~\cite{Krauss1997503}. The photon emitted from one nucleus can fluctuate to quark-antiquark pairs which then scatters off the other nucleus, emerging as a real vector meson. The photoproduction of vector meson is copious in relativistic heavy-ion collisions due to the large electric charge and the highly Lorentz-contracted electric field. In the production process, the elastic scattering occurs via Pomeron exchange, which impose a restriction on the production site within one of the two colliding nuclei. This indicates that the production source consists of two well separated nuclei (two slits). There are two possibilities: either nucleus 1 emits a photon, whereas nucleus 2 acts as a target or vice versa. The two possibilities are indistinguishable, which present the production process as a perfectly double-slit experiment of individual particle. Due to the negative parity of vector meson, the amplitudes from the two nuclei are assigned with opposite signs, which leads to a destructive interference. The destructive interference of $\rho^{0}$ for ultra-peripheral collisions (UPC) has been proposed by Klein and Nystrand~\cite{PhysRevLett.84.2330} and verified by the STAR Collaboration~\cite{PhysRevLett.102.112301}. In Refs.~\cite{PhysRevC.97.044910,PhysRevC.99.061901}, we extended the scenario to hadronic heavy-ion collisions (HHIC) and proposed to exploit the strong interactions in overlap region to test the observation effect on the interference.

The photons generated from the highly Lorentz-contacted electromagnetic field are expected to be fully linearly polarized. It has been suggested by C. Li et al.~\cite{LI2019576} that the collisions of the linearly polarized photons ($\gamma + \gamma \rightarrow l^{+} + l^{-}$) result in the 2nd-order and 4th-order modulations in azimuth (in the plane perpendicular to the beam direction) between the pair momentum and the lepton momentum for diletpon production. The experimental signature was confirmed by STAR Collaboration for the dielectron measurements~\cite{collaboration2019probing}. The vector meson production from linearly polarized photons also possesses distinctive signature in the asymmetries of the decay angular distributions. Continuous efforts~\cite{SCHILLING1968559,CRIEGEE1968282,PhysRevLett.24.1364,EISENBERG197661,DERRICK1996259,PhysRevC.77.034910,PhysRevC.97.035208} has been made for more than fifty years to utilize the linearly polarized photons in vector meson photoproduction as a parity filter for the exchange of particles in $t$ channel, which is an effective tool to separate the natural from unnatural parity exchange in $t$ channel. In this letter, we exploit the linearly polarized photons to investigate the interference effect for the vector meson photoproduction in heavy-ion collisions and demonstrate a periodic oscillation with transverse momentum for the asymmetries of the decay angular distributions. 

\begin{figure}[!htb]
	\includegraphics[width=1.0\hsize]
	{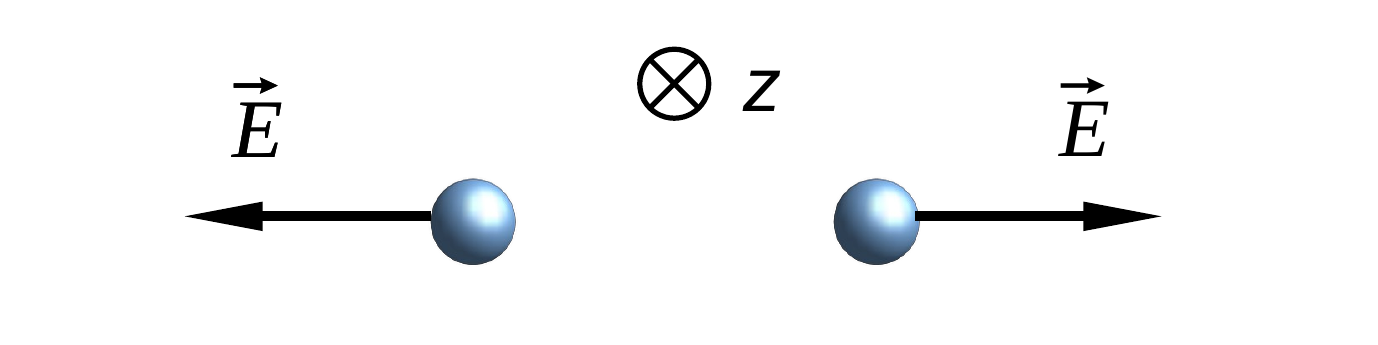}
	\caption{(Color online) Schematic diagram for the direction of electric vector in ultra-peripheral heavy-ion collisions.}
	\label{figure1}
\end{figure}

Hereinafter, we take the process of $\gamma + A \rightarrow \rho^{0} + A \rightarrow \pi^{+} + \pi^{-} +A$ in heavy-ion collisions to illustrate the interference effect on the asymmetries of the decay. The decay distribution is discussed in its helicity system: the $z$-direction is chosen to the direction of flight of vector meson in the photon-nucleon center of mass frame. The decay angles $\theta$ and $\phi$ are the polar and azimuthal angles of the decay daughters in the vector meson rest frame, respectively. In experiment, the direction of $z$-axis is approximated by that of incoming beams. It has been verified by Monte Carlo calculations that this is a good approximation. Here, we adopt the approximation in our calculation for direct comparisons with experimental results. Following Ref.~\cite{SCHILLING1970397}, the decay angular distribution of $\rho^{0}$ with the assumption of no helicity flip is:
\begin{equation}
\frac{d^{2}N}{d\cos\theta d\phi} =\frac{3}{8\pi}\sin^{2}\theta[1 + P_{\gamma}\cos2(\phi-\Phi)],
\label{equation1}
\end{equation}
where $P_{\gamma}$ is the degree of linear polarization, and $\Phi$ is the angle between the photon polarization plane and the $\rho^{0}$ production plane.  The helicity nonflip assumption is consistent with the spin density matrix element measurements for $\rho^{0}$ from STAR~\cite{PhysRevC.77.034910} and ZEUS~\cite{DERRICK1996259}. As revealed in Eq.~\ref{equation1}, the linearly polarized photons result in the 2nd-order modulations in azimuth between the vector meson momentum and the $\pi^{\pm}$ momentum. The magnitude of 2nd-order modulation gives: 
\begin{equation}
2\langle \cos(2\phi) \rangle = P_{\gamma}\cos(2\Phi).
\label{equation2}
\end{equation}
The strength of modulation relies on the degree and direction of linear polarization. The photons generated from the nuclei are fully linearly polarized, which gives $P_{\gamma} =1$. The orientation of linearly polarized photons is determined by the direction of the electric field vector. As demonstrated in Fig.~
\ref{figure1}, the electric vector for the photons, which interact with the nuclei, is parallel to the impact parameter vector. This suggests that the polarization plane is completely coincidence with the reaction plane. Here, we employ the point-like assumption for the nuclei, which is appropriate in ultra-peripheral collisions. We can redefine the $x$-axis as the impact parameter vector, then $\Phi$ is the azimuth angle of $\rho^{0}$ transverse momentum. For event by event average, the modulation strength can be written as:
\begin{equation}
2\langle \cos(2\phi) \rangle = P_{\gamma}\langle \cos(2\Phi) \rangle.
\label{equation3}
\end{equation}  
That is to say that the modulations of the decay distribution for $\rho^{0}$ are determined by the asymmetries in its two-dimensional transverse momentum distribution.

\begin{figure*}[!htb]
	\includegraphics[width=0.29\hsize]
	{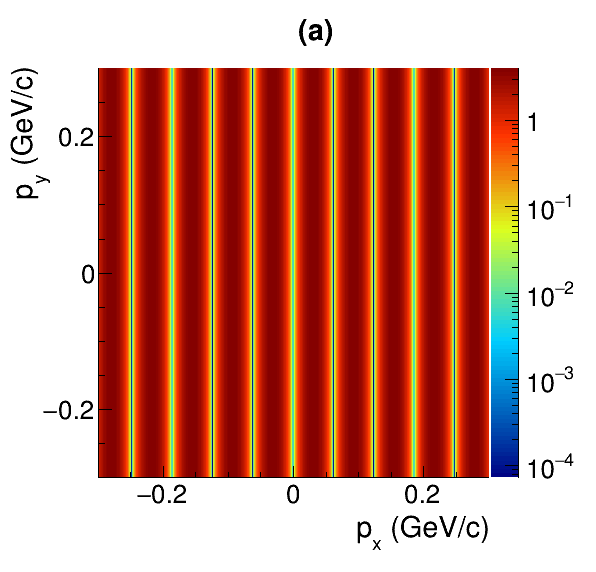}
	\includegraphics[width=0.32\hsize]
	{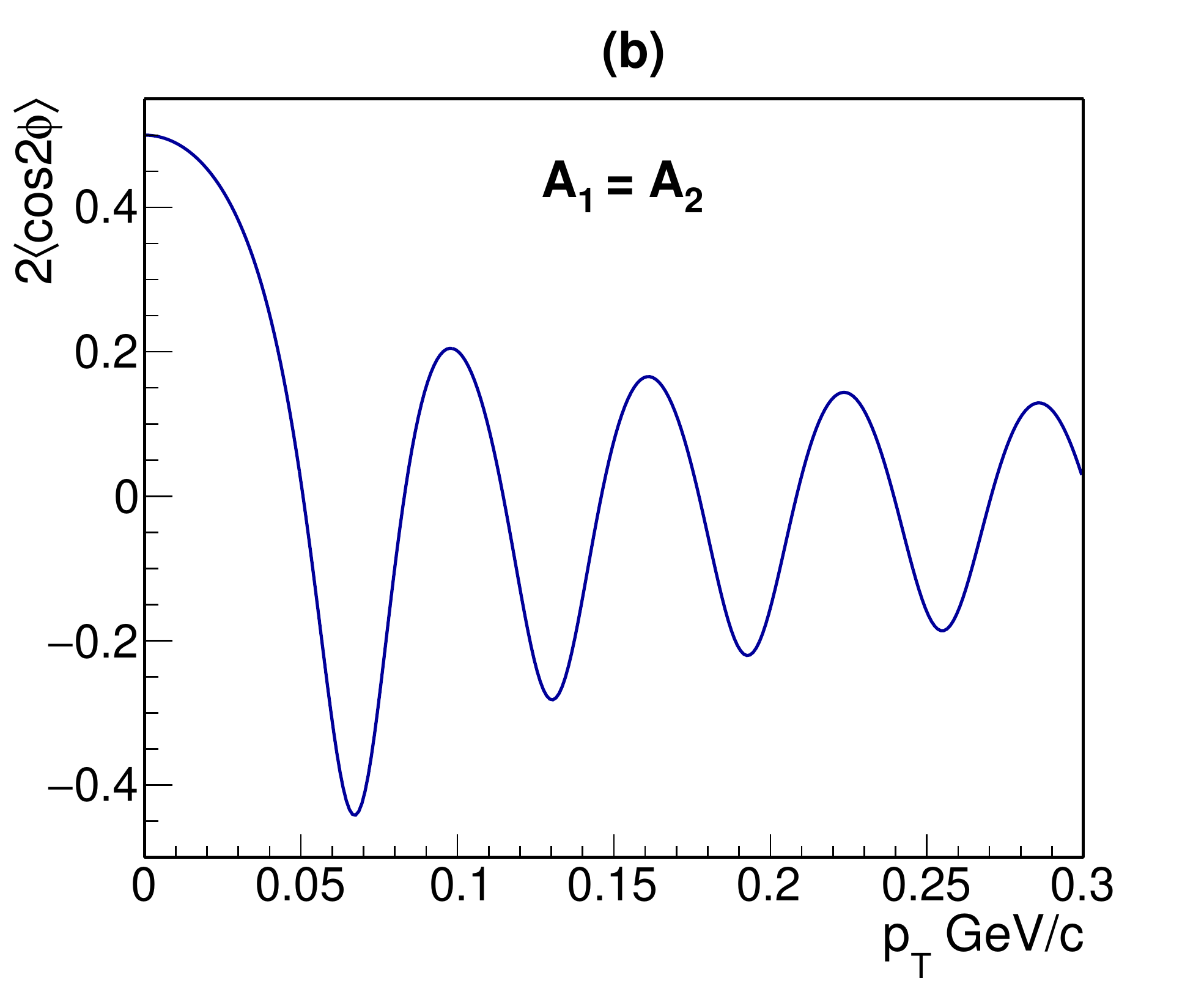}
	\includegraphics[width=0.32\hsize]
	{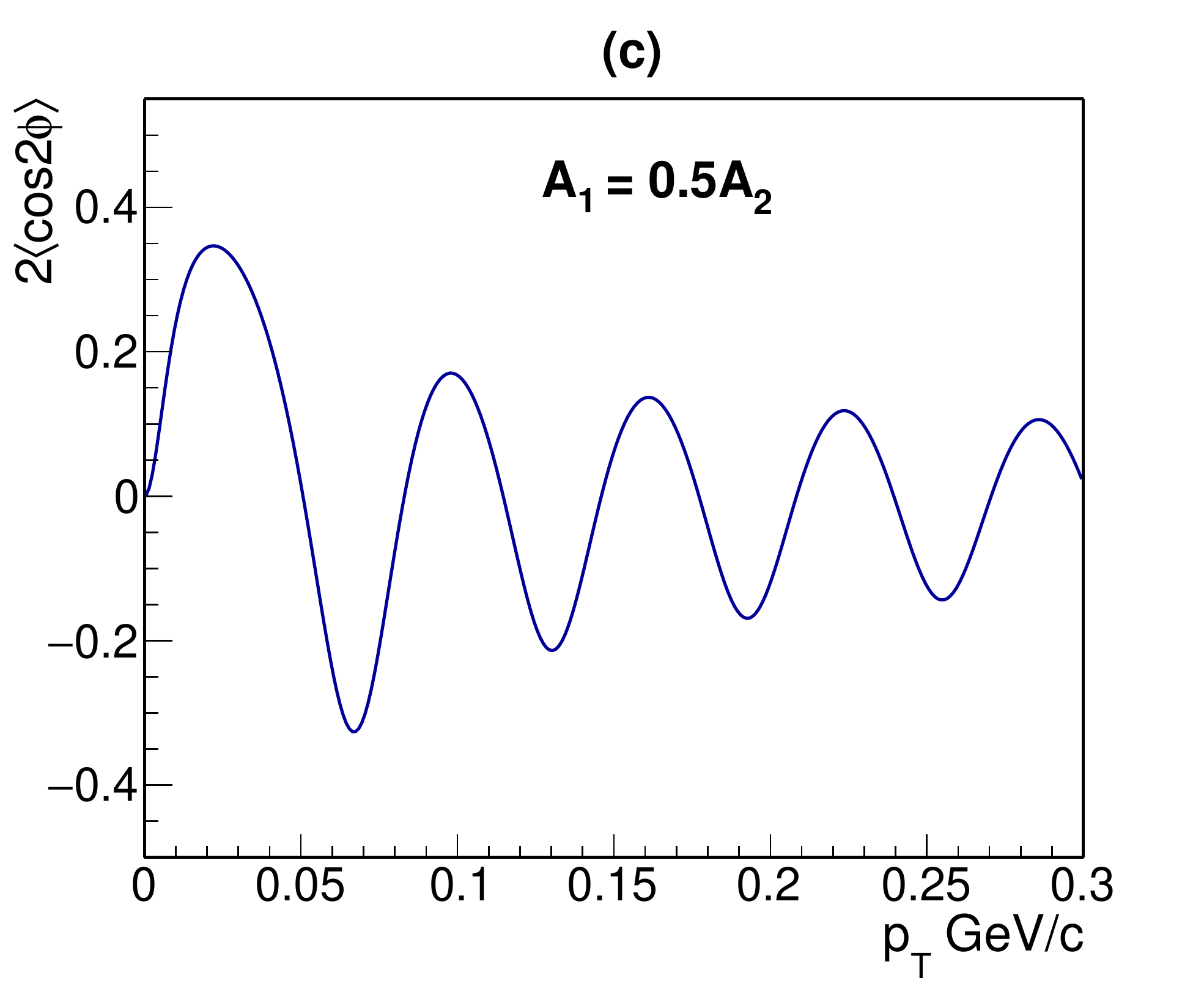}
	\caption{(Color online) Panel (a): two-dimensional transverse momentum distribution pattern with point-like assumption at mid-rapidity. Panel (b): the modulation strength, 2$\langle \cos\phi \rangle$, as a function of transverse momentum at mid-rapidity. Panel (c): 2$\langle \cos\phi \rangle$ as a function of transverse momentum at forward rapidity ($A_{1} \neq A_{2}$).}
	\label{figure2}
\end{figure*}

\begin{figure}[!htb]
	\includegraphics[width=1.0\hsize]
	{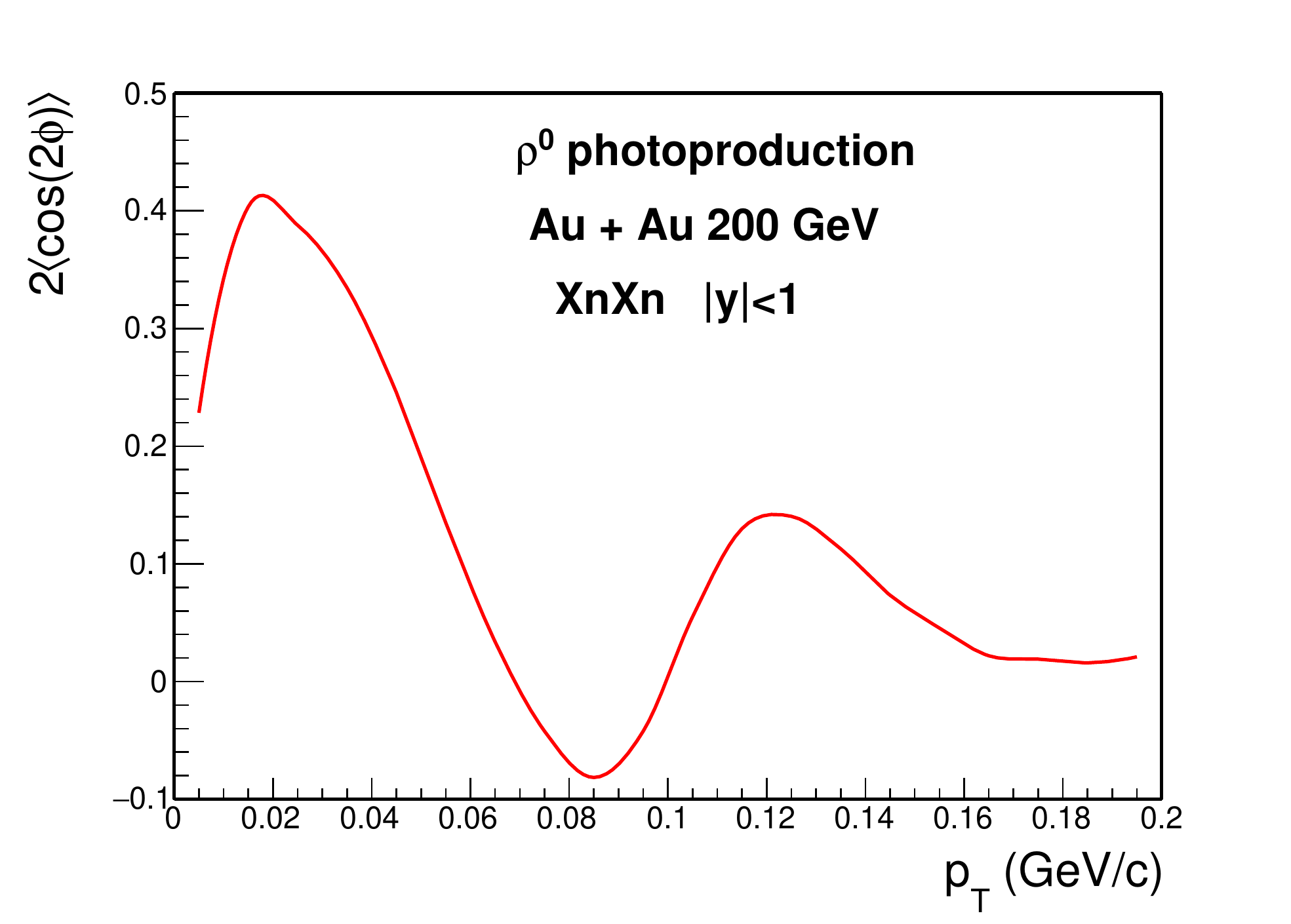}
	\caption{(Color online) The modulation strength, $2\langle \cos(2\phi) \rangle$, as a function of transverse momentum in ultra-peripheral Au+Au collisions at $\sqrt{s_{NN}} = 200$ GeV for mutual dissociation mode XnXn.}
	\label{figure3}
\end{figure}
  
  The probability distribution of $\rho^{0}$ photoproduciton in two-dimensional transverse momentum space can be expressed by performing a Fourier transformation to the amplitude in coordinate space:
  \begin{equation}
  \frac{d^{2}P}{dp_{x}dp_{y}} =\frac{1}{2\pi} |\int d^{2}x_{\perp} A(x_{\perp})e^{ip_{\perp} \cdot  x_{\perp}}|^{2},
  \label{equation4}
  \end{equation}
  where $A(x_{\perp})$ is the amplitude distribution in transverse plane. As a start, we can approximate the production over the two nuclei as two point sources. If there is no interference between the two source, the probability distribution in transverse momentum space will be uniform in azimuth, which would give no modulation for the decay distribution. With the interference effect, the amplitude distribution from the two colliding nuclei for point-like assumption can be written as:
    \begin{equation}
 A(x_{\perp}) = A_{1}\delta(x_{\perp} - \frac{1}{2}b) - A_{2}\delta(x_{\perp} + \frac{1}{2}b),
  \label{equation5}
  \end{equation}
 where $A_{1}$ and $A_{2}$ are the relative amplitudes for the two colliding nuclei, $\delta$ is the Dirac $\delta$ function, $b$ is the impact parameter, and the minus sign in the equation corresponds to the negative parity of $\rho^{0}$. The corresponding two-dimensional transverse momentum distribution with $b =$ 20 fm at mid-rapidity ($A_{1} = A_{2}$) is shown in Fig.~\ref{figure2} panel (a). The two-dimensional transverse momentum distribution reveals a typical Young’s double-slit interference pattern with a series of alternating light and dark fringes, which possesses significant asymmetries in azimuth. According to Eq.~\ref{equation3}, the modulation strength factor, $2 \langle \cos(2\phi) \rangle$, versus transverse momentum ($p_{T}$) can be extracted from the two-dimensional distribution and the result is shown in panel (b) of Fig.~\ref{figure2}. The modulation strength shows a periodic oscillation like the hyperbolic Bessel Functions with the transverse momentum. The oscillation cycle is determined by the distant between the two source (impact parameter).  For the production at forward rapidity in heavy-ion collisions, the relative amplitudes from the two nuclei are not equal. Panel (c) of Fig.~\ref{figure2} shows the modulation strength as a function of transverse momentum with unequal amplitudes from the two sources. Since the interference is not complete in this case, the oscillation level is weaker than that at mid-rapidity. Furthermore, the modulation strength is 0 at $p_{T} =$ 0, while it is maximum for the case at mid-rapidity. This is due to that the interference contribution is close to 0 for $p_{T} =$ 0.

As mentioned herein before, we have qualitatively demonstrated the interference effect on the asymmetries of the decay angular distributions for $\rho^{0}$ from photoproduction. The quantitative calculations is described herein after. Following Ref.~\cite{PhysRevC.97.044910}, the production amplitude distribution for $\rho^{0}$ photoproduction is determined by the spatial photon flux and the corresponding $\gamma  A$ scattering amplitude $\Gamma_{\gamma A \rightarrow \rho^{0} A}$. According to equivalent photon approximation, the spatial photon flux generated by the heavy nuclei can be written as:
\begin{equation}
\begin{aligned}
 \frac{d^{3}N_{\gamma}(\omega_{\gamma}, \vec{x}_{\perp})}{d\omega_{\gamma}d\vec{x}_{\perp}}& = \frac{4Z^{2}\alpha}{\omega_{\gamma}}|\int\frac{d^{2}\vec{k}_{\gamma\perp}}{(2\pi)^{2}}\vec{k}_{\gamma\perp} \frac{F_{\gamma}(\vec{k}_{\gamma})}{|\vec{k}_{\gamma}|^{2}}e^{i\vec{x}_{\perp}\cdot\vec{k}_{\gamma\perp}} |^{2}\\
& \vec{k}_{\gamma}=(\vec{k}_{\gamma\perp}, \frac{\omega_{\gamma}}{\gamma_{c}}), \omega_{\gamma}=\frac{1}{2}M_{\rho^{0}}e^{\pm y},
\end{aligned}
\label{equation6}
\end{equation}  
where $\vec{x}_{\perp}$ and $\vec{k}_{\gamma\perp}$ are two-dimensional photon position and momentum vectors perpendicular to the beam direction, $\omega_{\gamma}$ is the energy of emitted photon, $Z$ is the electric charge of nucleus, $\alpha$ is the electromagnetic coupling constant, $\gamma_{c}$ is the Lorentz factor of the beam, $M_{\rho^{0}}$ and $y$ are the mass and rapidity of $\rho^{0}$, and $F_{\gamma}(\vec{k}_{\gamma})$ is the nuclear electromagnetic form factor. The form factor can be obtained via the Fourier transformation of the charge density in the nucleus. The charge density profile of the nucleus can be parametrized by the Woods-Saxon distribution:
\begin{equation}
\rho_{A}(r) = \frac{a^{0}}{1+\rm{exp}[(r-R_{WS})/d]},
\label{equation7}
\end{equation}
where the radius $R_{WS}$ and skin depth $d$ are based on fits to electron-scattering data~\cite{WSD} and $a^{0}$ is the normalization factor.

The scattering amplitude $\Gamma_{\gamma A \rightarrow \rho^{0} A}$ with the shadowing effect can be estimated via Glauber\cite{doi:10.1146/annurev.nucl.57.090506.123020} + vector meson dominance (VMD) approach\cite{RevModPhys.50.261}:
\begin{equation}
\Gamma_{\gamma A \rightarrow \rho^{0} A}(\vec{x}_{\perp}) = \frac{f_{\gamma N \rightarrow \rho^{0}N}(0)}{\sigma_{\rho^{0}N}}[1-\rm{exp}(-\sigma_{\rho^{0}N}T(\vec{x}_{\perp}))],
\label{equation8}
\end{equation}
where $f_{\gamma N \rightarrow \rho^{0}N}(0)$ is the forward-scattering amplitude for $\gamma +  N \rightarrow \rho^{0} + N$, $\sigma_{\rho^{0} N}$ is the total $\rho^{0} N$ cross section, and $T(\vec{x}_{\perp})$ is the thickness function for nucleus defined as $T(\vec{x}_{\perp}) = \int_{-\infty}^{+\infty} \rho(\sqrt{\vec{x}^{2}_{\perp} + z^{2}})dz$. The $f_{\gamma N \rightarrow \rho^{0}N}(0)$ can be determined from the measurements of forward scattering cross section, $\frac{d\sigma_{\gamma N \rightarrow \rho^{0}N}}{dt}|_{t=0}$, which is well parametrized in Ref.~\cite{KLEIN2017258}. Using the optical theorem and VMD relation, the total cross for $\rho^{0} N$ scattering can be given by
\begin{equation}
\sigma_{\rho^{0}N} = \frac{f_{\rho^{0}}}{4\sqrt{\alpha}}f_{\gamma\rho^{0} \rightarrow \rho^{0}N},
\label{equation9}
\end{equation}
where $f_{\rho^{0}}$ is the $\rho^{0}$-photon coupling. To account for the coherent length effect, the thickness function $T(\vec{x}_{\perp})$ is redefined as
\begin{equation}
T(\vec{x}_{\perp}) = \int_{-\infty}^{+\infty} \rho(\sqrt{\vec{x}^{2}_{\perp} + z^{2}})e^{iq_{L}z}dz, q_{L} = \frac{M_{\rho^{0}}e^{y}}{2\gamma_{c}},
\label{equation10}
\end{equation}
where $q_{L}$ is the longitudinal momentum transfer required to produce a real $\rho^{0}$. Finally, the production amplitude distribution is given by  
\begin{equation}
A(\vec{x_{\perp}}) = \Gamma_{\gamma A \rightarrow \rho^{0} A} \sqrt{\frac{d^{2}N_{\gamma}}{d^{2}\vec{x_{\perp}}}}.
\label{equation11}
\end{equation}

In addition to the coherent production with interference, there are also incoherent contribution for $\rho^{0}$ photoproduction in heavy-ion collisions, which is significant at relative high transverse momentum. The incoherent production should have no contribution to the asymmetries of decay distribution, which would weaken the overall modulation strength. Similarly to the case of coherent $\rho^{0}$ photoproduction on nuclei, the incoherent cross section, $\sigma_{\gamma A \rightarrow\rho^{0} A^{'}}$, could be related to the exclusive $\gamma p $ cross section $\sigma_{\gamma p \rightarrow\rho^{0} p}$  via Glauber + VMD approach. Here, $A^{'}$ is the final nuclear state containing products of the nuclear disintegration.
The approach gives:
\begin{equation}
\begin{aligned}
\sigma_{\gamma A \rightarrow\rho^{0} A^{'}}= 
\sigma_{\gamma p\rightarrow\rho^{0} p} \int d^{2}{\vec{x}_{\perp}}T(\vec{x}_{\perp})e^{-\sigma^{in}_{\rho^{0} N}T(\vec{x}_{\perp})},
\end{aligned}
\label{equation12}
\end{equation}
where $\sigma^{in}_{\rho^{0} N}=\sigma_{\rho^{0} N}-\sigma^{2}_{\rho^{0} N}/(16\pi B_{\rho^{0}})$ is the inelastic $\rho^{0}$-nucleon cross section, and
B$_{\rho^{0}}$ is the slope of the $t$ dependence of the $\gamma p\rightarrow\rho^{0} p$ scattering\cite{KLEIN2017258}.

Here, we focus on the ultra-peripheral collisions, where there is no nuclear overlap to reject hadronic background. According to the optical Glauber model~\cite{Miller:2007ri}, the mean number of projectile nucleons that interact at least once in $A+A$ collisions with impact parameter $b$ is:
\begin{equation}
P_{H}(b) = \int d^{2}\vec{x_{\perp}} T(\vec{x_{\perp}} - \vec{b}) \{ 1 - exp[-\sigma_{NN}T(\vec{x_{\perp}})]\},
\label{equation13}
\end{equation}
where $\sigma_{NN}$ is the total nucleon-nucleon cross section. Then, the probability of having no hadronic interaction is $exp[-P_{H}(b)]$.  In experiment, the UPC events are usually triggered or selected by the mutual Coulomb dissociation of the nuclei, in which the number of emitted forward neutrons can be detected by Zero Degree Calorimeters. Following the EPA approach, the Coulomb excitation probability of an ultra-relativistic nucleus can be determined by the photon flux companied with the nucleus and the appropriate photon-absorption cross section of nuclei. The lowest-order probability for an excitation of nucleus which emits at least one neutron ($X_{n}$) is
\begin{equation}
\label{equation14}
m_{Xn}(b) = \int dk n(b,E) \sigma_{\gamma A\rightarrow A^{*}}(E),
\end{equation}
where $E$ is the photon energy,b is the impact parameter, $n(b,E)$ is photon flux described by Eq.~\ref{equation6}, and $\sigma_{\gamma A\rightarrow A^{*}}(E)$ is the photoexcitation cross section with incident energy $E$. The photoexcitation cross section $\sigma_{\gamma A\rightarrow A^{*}}(E)$ can be extracted from the experimental measurements~\cite{VEYSSIERE1970561,LEPRETRE1981237,CARLOS1984573,PhysRevD.5.1640,PhysRevD.7.1362,PhysRevLett.39.737,ARMSTRONG1972445}. The probability of mutual dissociation for the two nuclei with at least one neutron emission for each beam (XnXn)is then given by:
\begin{equation}
\label{equation14}
P_{XnXn}(b) = (1-\rm{exp}[-m_{Xn}(b)])^{2}.
\end{equation}
As described in our recent work~\cite{br2020acoplanarity}, the mutual dissociation probability with any number of neutron emission can be estimated in similar way.

With the inputs described herein before,  the cross section and differential distributions for $\rho^{0}$ photoproduction
can be calculated, which could reasonably described the measurements from STAR~\cite{PhysRevC.77.034910,PhysRevC.96.054904} and ALICE~\cite{articleJHEP,collaboration2020coherent}. The estimated modulation strength, $2\langle \cos(2\phi) \rangle$, of $\rho^{0}$ as a function of transverse momentum in ultra-peripheral Au+Au collisions at $\sqrt{s_{NN}} = 200$ GeV for mutual dissociation mode XnXn is shown in Fig.~\ref{figure3}. Similar to the case for point-like assumption, the modulation strength shows a periodic oscillation with the transverse momentum. The amplitude of the second oscillation period is weaker than that of the first oscillation period, which is due to more fraction of incoherent production at higher $p_{T}$. The oscillation feature disappears for $p_{T} >$ 0.16 GeV/c, where the incoherent production plays dominate role in the photoproduction. The results from calculations can reasonably describe the transverse momentum dependence of the 2nd-order modulation in azimuth for the $\rho^{0}$ decay observed by the STAR collaboration~\cite{DanielQM}. 

In Summary, the double slit interference is explored in polarization dimension with the linearly polarized photons accompanied by relativistic nuclei in heavy-ion collisions. We demonstrate how the interference between the two colliding nuclei affect the asymmetries of the decay angular distributions for vector meson photoproduction from linearly polarized photons. Using the vector meson dominance with Glauber approach, the 2nd-order modulation in azimuth for the $\rho^{0}$ decay from photoproduction in ultra-peripheral Au+Au collision at $\sqrt{s_{NN}} =$ 200 GeV is estimated, and reveals a periodic oscillation with the transverse momentum, which could reasonably described the decay asymmetries observed by the STAR Collaboration. The calculation can be easily extended to other vector mesons, e.g. $\omega$ and $\phi$, which could be studied by future experiments.

  This work was funded by the National Natural Science Foundation of China under Grant Nos. 11775213, 11505180 and 11375172, the U.S. DOE Office of Science under contract No. DE-SC0012704, and MOST under Grant No. 2014CB845400.

\nocite{*}
\bibliographystyle{aipnum4-1}
\bibliography{aps}
\end{document}